\documentclass[12pt]{article}
\usepackage{a4wide,epsfig,psfrag,amsmath,amssymb,cite,scalefnt}
\usepackage{color}

\graphicspath{ {figs/} }

\newcommand{\be}{\begin{equation}}
\newcommand{\ee}{\end{equation}}
\newcommand{\bea}{\begin{eqnarray}}
\newcommand{\eea}{\end{eqnarray}}

\def\LO{\ifmmode \textrm{LO} \else LO\fi}
\def\NLO{\ifmmode \textrm{NLO} \else NLO\fi}
\def\NNLO{\ifmmode \textrm{NNLO} \else NNLO\fi}

\allowdisplaybreaks

\newcommand{\gsim}{\;\rlap{\lower 3.5 pt \hbox{$\mathchar \sim$}} \raise 1pt
 \hbox {$>$}\;}
\newcommand{\lsim}{\;\rlap{\lower 3.5 pt \hbox{$\mathchar \sim$}} \raise 1pt
 \hbox {$<$}\;}

\begin{document}

\title{\vskip-3cm{\baselineskip14pt
    \begin{flushleft}
      \normalsize DESY 15-133, TTP15-025
  \end{flushleft}}
  \vskip1.5cm
  Light-by-light-type corrections to the muon anomalous magnetic moment at four-loop order
}

\author{
  Alexander~Kurz$^{a,b}$,
  Tao~Liu$^a$,
  Peter~Marquard$^b$,
  Alexander V.~Smirnov$^c$,\\
  Vladimir A.~Smirnov$^d$,
  Matthias Steinhauser$^a$
  \\[1em]
  {\small\it $(a)$ Institut f{\"u}r Theoretische Teilchenphysik, 
    Karlsruhe Institute of Technology (KIT),}\\
  {\small\it 76128 Karlsruhe, Germany}
  \\
  {\small\it $(b)$ Deutsches Elektronen Synchrotron DESY, Platanenallee 6}\\
  {\small\it 15738 Zeuthen, Germany}
  \\
  {\small\it $(c)$ Scientific Research Computing Center, Moscow State University,}\\
  {\small\it 119991, Moscow, Russia}
  \\
  {\small\it $(d)$ Skobeltsyn Institute of Nuclear Physics of Moscow State University,}\\
  {\small\it 119991, Moscow, Russia}
}

\date{}

\maketitle

\thispagestyle{empty}

\begin{abstract}
  The numerically dominant QED contributions to the anomalous magnetic moment
  of the muon stem from Feynman diagrams with internal electron loops. We
  consider such corrections and present a calculation of the four-loop
  light-by-light-type corrections where the external photon couples to a
  closed electron or muon loop.  We perform an asymptotic
  expansion in the ratio of electron and muon mass and reduce the resulting
  integrals to master integrals which we evaluate using analytical and numerical
  methods.  We confirm the results present in the literature which are based
  on different computational methods.

  \medskip

  \noindent
  PACS numbers: 12.20.-m 12.38.Bx 14.60.Ef

\end{abstract}

\thispagestyle{empty}


\newpage


\section{Introduction}

The anomalous magnetic moment of the muon provides an important test of the
Standard Model of particle physics. It has been measured to an impressive accuracy
at the Brookhaven National Laboratory~\cite{Bennett:2006fi,Roberts:2010cj}
and two new experiments at Fermilab~\cite{Carey:2009zzb} and J-PARC~\cite{Mibe:2010zz}
are planned to further improve the measured value. 

On the theory side much effort has been made to provide a precise prediction
for the muon magnetic moment, see
e.g. Refs.~\cite{Melnikov:2006sr,Jegerlehner:2009ry,Miller:2012opa} for
detailed reviews. The comparison of the precise measurements and calculations
shows a deviation of about three standard deviations, which already persists
for several years. This fact makes the anomalous magnetic moment of the
muon, $a_\mu$, an interesting quantity for further investigations.

Several ingredients are needed to obtain the theory prediction for $a_\mu$.
The numerically most important one origins from QED radiative corrections
which are analytically known up to three
loops~\cite{Laporta:1992pa,Laporta:1993ju,Laporta:1996mq} and
numerically up to five-loop order~\cite{Aoyama:2012wk}. Also the electroweak
correction, which are known at the two-loop level, are under
control~\cite{Czarnecki:1995sz,Knecht:2002hr,Czarnecki:2002nt,Gnendiger:2013pva}.
The dominant contribution to the uncertainty comes from the hadronic
contribution which can be split into a vacuum polarization and light-by-light
contribution. The vacuum polarization contribution is obtained with the help
of a dispersion integral over the experimentally measured cross section
$e^+e^-\to\mbox{hadrons}$ where the dominant contribution comes from low
energies. The corresponding analysis has been performed at leading
order~\cite{Davier:2010nc,Hagiwara:2011af,Jegerlehner:2011ti,Benayoun:2012wc},
next-to-leading
order~\cite{Krause:1996rf,Greynat:2012ww,Hagiwara:2003da,Hagiwara:2011af} and
next-to-next-to-leading order~\cite{Kurz:2014wya}.  The least known
contribution origins from the hadronic light-by-light part which has been
considered by several groups at leading
order~\cite{Nyffeler:2009tw,Melnikov:2003xd,Bijnens:2007pz}.  The
corresponding next-to-leading order effects have been estimated to be
small~\cite{Colangelo:2014qya}.

\begin{figure}[t]
  \begin{center}
    \mbox{}\hfill
    \includegraphics[scale=0.8]{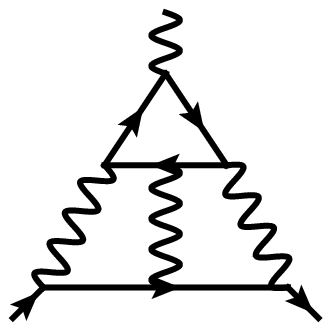}
    \hfill
    \includegraphics[scale=0.8]{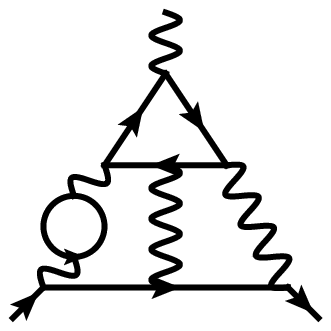}
    \hfill
    \includegraphics[scale=0.8]{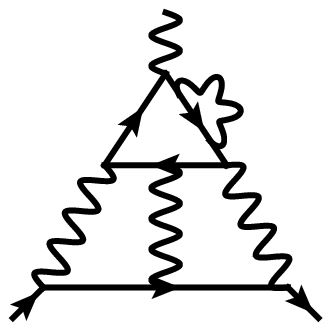}
    \hfill
    \includegraphics[scale=0.8]{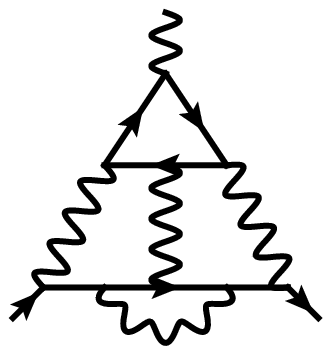}
    \hfill\mbox{}
    \\[1em]
    \mbox{} \hspace*{4em}
    3 loops \hfill IV(a) \hfill IV(b) \hfill IV(c)
    \hspace*{4em} \mbox{}
  \end{center}
  \caption{\label{fig::FDs}Sample light-by-light-type Feynman diagrams
    contributing to $a_\mu$. The external solid line represents the muon and at least
    one of the internal solid loops denotes electrons. In the case of IV(a) the
    second fermion loop can either be an electron or muon loop. Wavy lines
    represent photons.}
\end{figure}

In this paper we focus on the QED contribution to the muon
anomalous magnetic moment which can be cast in the form
\begin{eqnarray}
  a_\mu &=& \sum_{n=1}^\infty a_\mu^{(2n)} \left( \frac{\alpha}{\pi} \right)^n
  \,,
  \label{eq::amu}
\end{eqnarray}
where $\alpha$ is the fine structure constant. The first three coefficients 
on the right-hand-side, which correspond to the one-, two- and three-loop
corrections, are known
analytically~\cite{Schwinger:1948iu,Petermann:1957hs,Sommerfield:1957zz,Elend:1966,Samuel:1990qf,Laporta:1992pa,Laporta:1993ju,Laporta:1996mq,Czarnecki:1998rc,Passera:2006gc}. For
the four- and five-loop contributions only numerical
results are
available~\cite{Kinoshita:2004wi,Aoyama:2007mn,Aoyama:2012wk}. Note, that
even for the four-loop coefficient $a_\mu^{(8)}$ there
is no systematic cross check by an independent calculation; only a few
special cases have been computed analytically (see, e.g.,
Refs.~\cite{Laporta:1993ds,Aguilar:2008qj,Lee:2013sx,Baikov:2012rr}). 
An independent calculation of $a_\mu^{(8)}$ is
important since the four-loop contribution in
Eq.~(\ref{eq::amu}) amounts to\footnote{The ellipses stand for further digits
  and small contributions which are not shown.}
\begin{eqnarray}
  \left(-1.910\ldots + 132.685\ldots|_e + \ldots\right) \left( \frac{\alpha}{\pi} \right)^4
  \approx  381 \times 10^{-11}\,,
  \label{eq::amu_4l}
\end{eqnarray}
which is comparable to the deviation between the experimentally measured and 
theoretically predicted result for $a_\mu$ 
given by~\cite{Aoyama:2012wk}
\begin{eqnarray}
  a_\mu({\rm exp}) - a_\mu({\rm SM}) &\approx& 249(87) \times 10^{-11}
  \,.
  \label{eq::amu_diff}
\end{eqnarray}
In Eq.~(\ref{eq::amu_4l}) we have separated the contributions containing at least
one closed electron loop (second term on left-hand-side) from the pure
photonic part. The former are numerically dominant\footnote{This is also true
  at two and three loops, see. e.g., Ref.~\cite{Jegerlehner:2009ry} for explicit
  results.} which provides the motivation to concentrate in a first step on
these contributions. Note that about 95\% of the electron loop
contribution originates from the so-called light-by-light-type Feynman
diagrams where the external photon couples to a closed fermion
loop. Such contributions arise for the first time at three loops; see
Fig.~\ref{fig::FDs} for sample diagrams.  In this paper we perform an
independent calculation of the four-loop corrections.

It is convenient to decompose the four-loop term $a_\mu^{(8)}$
into a purely photonic piece and contributions involving electron and/or $\tau$
loops. Following Ref.~\cite{Aoyama:2012wk} we write
\begin{eqnarray}
  a_\mu^{(8)} &=& A_1^{(8)} + A_2^{(8)}(m_\mu / m_e) + A_2^{(8)}(m_\mu / m_\tau) 
  + A_3^{(8)}(m_\mu / m_e, m_\mu / m_\tau)
  \,.
\end{eqnarray}
$A_2^{(8)}(m_\mu / m_\tau)$ has been computed in Ref.~\cite{Kurz:2013exa}
using an asymptotic expansion for $m_\mu^2 \ll m_\tau^2$. Analytic results have
been obtained for several expansion terms which show a rapid convergence.
$A_2^{(8)}(m_\mu / m_\tau)$ and $A_3^{(8)}(m_\mu / m_e, m_\mu / m_\tau)$ 
are suppressed by $m_\mu^2/m_\tau^2$ and thus they are
numerically small.

$A_2^{(8)}(m_\mu / m_e)$ can be split into
light-by-light-type contributions (cf. Fig.~\ref{fig::FDs}) and contributions
where the external photon couples to the external muon line.  The leading term
of the latter can be obtained from calculations where in a first step the
electron mass is set to zero and the fine structure constant is renormalized
in the $\overline{\rm MS}$ scheme. Afterwards $\alpha$ is transformed to the
on-shell scheme which introduces $\log(m_e/m_\mu)$ terms in the final
result. Using this approach, the non-light-by-light contributions with two
closed electron loops have been computed analytically in
Ref.~\cite{Lee:2013sx}.

In this paper we compute the four-loop light-by-light contributions to
$A_2^{(8)}(m_\mu / m_e)$ which are exemplified by three Feynman diagrams in
Fig.~\ref{fig::FDs}. In case the external photon couples to a closed electron
loop it is not possible to set $m_e=0$ since this generates infrared
singularities. To circumvent this problem we perform an asymptotic expansion
for
$m_e\ll m_\mu$ which is described in some detail in Section~\ref{sec::calc}.
Results for contribution IV(a) with two closed electron loops have already
been considered 40 years ago in Refs.~\cite{Calmet:1975tw,Chlouber:1977dr}. 
In Section~\ref{sec::res} we will discuss in detail our results and 
compare to their findings and also to the ones in Ref.~\cite{Aoyama:2012wk}.
Section~\ref{sec::concl} contains our conclusions.


\section{\label{sec::calc}Technical details}

The Feynman integrals which contribute to the light-by-light part
of $A_2^{(8)}(m_\mu / m_e)$ contain two widely separated scales, which 
provides a small expansion parameter
\begin{eqnarray}
  x &=& \frac{m_e}{m_\mu} \,\, \approx \,\, 1/206.7682843
  \,.
\end{eqnarray}
Thus, it can be expected that already a few expansion terms provide a good
approximation to the exact result.  We compute four terms and show that the
one of order $x^3$ leads to negligible contributions. The linear and quadratic
term, however, can still lead to sizable contributions since the coefficients
of $x^n$ contain $\log(x)\approx -5.3$ terms which at four-loop order are
raised up to forth power.

We have implemented the asymptotic expansion using two different
programs. 
In the first approach we use the Mathematica package {\rm
  asy}~\cite{Pak:2010pt,Jantzen:2012mw} which is based on expansion by
regions~\cite{Beneke:1997zp,Smirnov:2002pj} formulated at the level of the
alpha representation~\cite{Smirnov:1999bza}. It provides the possibility to
obtain the asymptotic behaviour of a Feynman diagram in a given limit.
In fact, the output of
{\tt asy} are scaling rules for the alpha parameters. To exploit this
information one has to find a distribution of the external momentum and the 
loop momenta obeying certain scaling rules which we obtain by trying out all
possible combinations.

The second approach is based on an in-house program which generates all
possible combinations of loop momenta and external momentum and assigns 
for each combination all possible scalings of the loop momenta
(i.e. each loop momentum can either be soft or hard). In this way
one obtains  by construction all contributing regions. However, a double
counting is introduced since the
routing of the loop momenta is not unique. The double counting is eliminated
with the help of the unique alpha representation which is generated for each
momentum distribution.

We have applied both methods to each Feynman diagram and have obtained
identical final results. Note, however, that {\tt asy} requires significantly
more CPU time than our in-house program which is tailored for the problem at hand.

\begin{figure}[t]
  \centering
  \includegraphics[width=\textwidth]{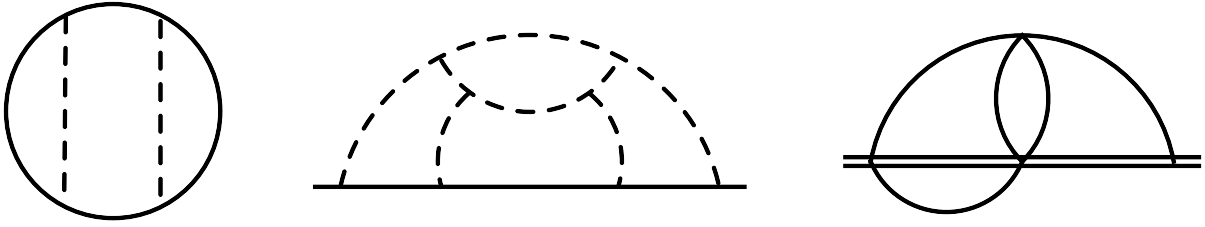}
  \caption{\label{fig:s1}Sample Feynman diagrams for the three appearing integral classes
    introduced in Eqs.~(\ref{eq::int_vac}) (left),~(\ref{eq::int_os}) (middle)
    and~(\ref{eq::int_lin}) (right). Solid and dashed lines denote massive and
    massless propagators. Double lines stand for denominators of the form
    $1/(2k\cdot q)$.}
\end{figure}

The output of the asymptotic expansion is manipulated with {\tt
  FORM}~\cite{Vermaseren:2000nd,Kuipers:2012rf} and 
{\tt TFORM}~\cite{Tentyukov:2007mu} (see also Ref.~\cite{Steinhauser:2015wqa})
which are used to perform traces and to deal with tensor structures. Afterwards
only scalar integrals are left which are reduced to master integrals with the
help of {\tt FIRE}~\cite{Smirnov:2014hma} and {\tt crusher}~\cite{crusher}.
For the classification of the master integrals we introduce the following set
of propagators
\begin{equation}
  \label{eq:1}
  P_i = \left \{ \frac{1}{k^2},\frac{1}{k^2 - M^2}, \frac{1}{k^2 - 2 k \cdot
      q},\frac{1}{2 k \cdot q}  \right \}
  \,,\qquad M\in\{m_e,m_\mu\}\,,
\end{equation}
where $k$ is a linear combination of loop momenta and
$q$, the external momentum,
with $q^2=m_\mu^2$.  Using these propagators we can build the required
three integral classes
\begin{description}
\item[Vacuum integrals] 
  \begin{equation}
    V(i_1,...,i_n) = \int \left ( \prod_{j=1}^L d^D k_j \right ) D_1^{i_1}\cdots
    D_n^{i_n},\, \qquad D_i\in\{P_1,P_2\} \,,
    \label{eq::int_vac}
  \end{equation}
\item[On-shell integrals] 
  \begin{equation}
    O(i_1,...,i_n) = \int \left ( \prod_{j=1}^L d^D k_j \right ) D_1^{i_1}\cdots
    D_n^{i_n} ,\, \qquad D_i\in\{P_1,P_2,P_3\} \,,
    \label{eq::int_os}
  \end{equation}
\item[Linear integrals] 
  \begin{equation}
    L(i_1,...,i_n) = \int \left ( \prod_{j=1}^L d^D k_j \right ) D_1^{i_1}\cdots
    D_n^{i_n} ,\,\qquad  D_i\in\{P_1,P_2,P_4\} \,,
    \label{eq::int_lin}
  \end{equation}
\end{description}
which are exemplified in Fig.~\ref{fig:s1}. 

To get an impression how the individual types of integrals arise after
asymptotic expansion we discuss in some detail the three-loop case (cf. left
diagram in Fig.~\ref{fig::FDs}). The valid regions are obtained by considering
appropriate routings of the loop and external momenta, allowing
each loop momentum to be either soft ($k\sim m_e$)
or hard ($k\sim m_\mu$). In total we obtain eight possible regions.
In case all loop momenta are hard the electron propagators are expanded
in $m_e$ and one ends up with three-loop on-shell integrals. On the other
hand, in case all loop momenta are soft the muon propagators are expanded
for $k^2\ll 2k\cdot q$ and one has
\begin{eqnarray}
  \frac{1}{m_\mu^2-(k-q)^2} &=& \frac{1}{2k\cdot q-k^2} \,\,=\,\,
  \frac{1}{2k\cdot q}\left(1 + \frac{k^2}{2k\cdot q} + \ldots\right)
  \,,
\end{eqnarray}
which leads to linear integrals where $m_e$ sets the mass scale.  If two loop
momenta are hard and one is soft one obtains two-loop on-shell (with mass
scale $m_\mu$) and one-loop vacuum integrals (with mass scale $m_e$).  The
remaining three regions, i.e. two soft and one hard loop momentum, leads to
one-loop on-shell integrals and either vacuum or linear integrals where the
massive scale is given by the electron mass.  It is interesting to note that
on-shell and vacuum integrals only occur for the even powers of $x$ whereas
linear integrals are present both for even and odd powers.  Note,
however, that the
corresponding master integrals differ: the $x^0$ and $x^2$ terms
involve master integrals with even number of linear propagators whereas for
the $x^1$ and $x^3$ terms their number is odd.

The pattern observed at three loops repeats itself at four loops. 
For some diagrams one obtains more than 20 regions 
leading to single-scale integrals of the type
\begin{itemize}
\item four-loop on-shell,
\item four-loop linear,
\item products of three-loop on-shell and one-loop vacuum,
\item products of two-loop on-shell and two-loop vacuum or linear,
\item products of one-loop on-shell and three-loop vacuum or linear,
\item and products two one-loop and one two-loop integrals involving 
  vacuum, on-shell and/or linear integrals.
\end{itemize}
The mass scale of the on-shell integrals is given by $m_\mu$ and the
one of the vacuum and linear integrals by $m_e$, with the following exception:
one-loop vacuum integrals with mass scale $m_\mu$ occur in IV(a)
in case a closed muon loop is present.

Massive vacuum integrals, which are only needed up to three loops, are well
documented in the literature~\cite{Avdeev:1995eu,Steinhauser:2000ry}. 
All one-, two- and three-loop on-shell integrals entering our calculation are known
analytically~\cite{Lee:2010ik}. There are about 70 four-loop on-shell integral 
which are needed for IV(a), IV(b) and IV(c), about 40 are known
analytically or to high numerical precision. All of them  
are taken from the calculation of the 
four-loop $\overline{\rm MS}$-on-shell quark mass relation performed in
Ref.~\cite{Marquard:2015qpa}. 
We furthermore require about 70 four-loop linear master integrals 
where about 20 have been computed analytically or to high numerical precision
using Mellin-Barnes techniques~\cite{Czakon:2005rk}. The remaining ones
have been computed using the package {\tt FIESTA}~\cite{Smirnov:2013eza}.
We propagate the uncertainty from each $\epsilon$ coefficient of each
master integral to the final result. To obtain the final error estimate we add 
the uncertainties in quadrature.

The renormalization of the light-by-light contribution only involves one-loop
on-shell counterterms for the fine structure constant, the muon and electron
masses and the wave function which are well established in the literature
(see, e.g., the textbook~\cite{Ryder:1985wq}).


\newcommand{\xdummy}{}
\newcommand{\x}     {x}
\newcommand{\xtwo}  {x^2}
\newcommand{\xthree}{x^3}
\newcommand{\xfour} {x^4}
\newcommand{\lx}     {\ell_x}
\newcommand{\lxtwo}  {\ell_x^2}
\newcommand{\lxthree}{\ell_x^3}
\newcommand{\lxfour} {\ell_x^4}
\newcommand{\XX}     {}
\newcommand{\XXtwo}  {}
\newcommand{\XXthree}{}
\newcommand{\XXfour} {}
\newcommand{\XLtwo}     {}
\newcommand{\XLtwotwo}  {}
\newcommand{\XLtwothree}{}
\newcommand{\XLtwofour} {}

\section{\label{sec::res}Discussion of results}

We start with looking at the three-loop light-by-light contribution to
$A_2^{(6)}(m_\mu / m_e)$ which is known analytically~\cite{Laporta:1992pa}.  
Using the method described in the previous section we obtain in numerical form 
($x=m_e/m_\mu$)
\begin{eqnarray}
  A_{2, \rm lbl}^{(6)}(m_\mu / m_e) &=&
   \xdummy  -14.5525 
  -6.5797  \lx
  \nonumber\\&&\mbox{}
  + \x \left[  18.0169 
  -13.1595  \lx
  \right]
  \nonumber\\&&\mbox{}
  + \xtwo \left[ -12.4582 
  +  1.5944  \lx
  -5.5700  \lxtwo
  +  0.6667  \lxthree
  \right]
  \nonumber\\&&\mbox{}
  + \xthree \left[ -12.0628
  -10.9662  \lx
  \right]
  \nonumber\\&&\mbox{}
  + \xfour \left[  14.4529 
  -19.0431  \lx
  +  2.2778  \lxtwo
  -0.7778  \lxthree
  \right]
  \nonumber\\&=&
   -14.5525 
  +  35.0805  \XLtwo
  \nonumber\\&&\mbox{}
  + \XX \left[  0.08714 
  +  0.3393  \XLtwo
  \right]
  \nonumber\\&&\mbox{}
  + \XXtwo \left[ -0.0002914 
  -0.0001988  \XLtwo
  -0.003703  \XLtwotwo
  -0.002363  \XLtwothree
  \right]
  \nonumber\\&&\mbox{}
  + \XXthree \left[ -0.0000014
  + 0.0000066 \XLtwo
  \right]
  \nonumber\\&&\mbox{}
  + \XXfour \left[ 0.0000000079
  + 0.000000056  \XLtwo
  + 0.000000035  \XLtwotwo
  + 0.000000064 \XLtwothree
  \right]
  \nonumber\\&=&
   \xdummy 20.5280 
  + \XX 0.4265 
  -0.006557 + 0.0000052+  0.0000002
  \nonumber\\&=&
   \xdummy 20.9479\,,
  \label{eq::a6}
\end{eqnarray}
where terms of order $x^5$ are neglected and 
$\ell_x = \log(x) \approx -5.3$.  
For completeness we provide analytic results in the Appendix.
One observes that odd powers in $x$
contain at most linear terms in $\ell_x$ whereas even powers contain terms up to
$\ell_x^3$.  After the second equality sign the numerical value for $x$ is
inserted, however, the contribution from the various powers of $x$ and
$\ell_x$ are kept separately.  At order $x^0$ and $x^1$ the logarithmic
contribution dominates over the constant. At order $x^2$ the constant and
linear logarithmic term is of the same order of magnitude and about a factor
ten smaller than the quadratic and cubic contribution. 
After the third
equality sign all contributions to $x^n$ are added. One observes that the
overall contribution of the $x^2$ term is already quite small and amounts to
only 0.03\% of the leading term. The linear term provides a 2\% contribution
and is still important. Let us mention that the cubic and quartic terms are
below 0.00003\% and are thus negligible.  
For completeness we present the final result for
$A_{2,\rm lbl}^{(6)}(m_\mu / m_e)$ after the last equality sign.

We now turn to the four-loop results.  For convenience we split IV(a) into
three contributions: IV(a0) contains two closed electron loops, in IV(a1) only
the fermion loop with the coupling to the external photon contains electrons,
and in IV(a2) electrons are running only in the two-point polarization
function and muons in the other fermion loop.  Let us mention that the
coefficients of the logarithmic contributions in the case of IV(a0) are known
analytically since the contributing four-loop on-shell integrals are available
in the literature~\cite{Lee:2013sx} and only the four-loop linear integrals
have to be evaluated numerically. As a consequence, it is possible to
reconstruct the pole terms of the IV(a0) contribution, which are in one-to-one
correspondence to the coefficients of the logarithms, analytically.
Similar arguments can be used to obtain the logarithmic contributions of
IV(a1) and IV(a2), for the $x^1$ and $x^3$ terms of IV(b) and for the 
$x^0$, $x^1$ and $x^3$ terms of IV(c). They are given in the Appendix. In
the main part of the paper we restrict ourselves to numerical results.

Using the same scheme for presenting the results as in Eq.~(\ref{eq::a6}) we
obtain for the five four-loop light-by-light-type contributions 
\begin{eqnarray}
  A_2^{(8),\rm IV(a0)} &=&
   7.5018 \pm  0.0026
  +  14.8808 \lx
  +   6.5797 \lxtwo
  \nonumber\\&&\mbox{}
  + \x \left[  6.29 \pm  0.46
  - 14.6216  \lx
  +  8.7729  \lxtwo
  \right]
  \nonumber\\&&\mbox{}
  + \xtwo \left[ -16.81 \pm  0.43
  +  30.0172 \lx
  -6.5069  \lxtwo
  + 7.6489  \lxthree
  -0.8889  \lxfour
  \right]
  \nonumber\\&&\mbox{}
  + \xthree \left[ -48.31 \pm  0.24
  -4.8739 \lx
  +  13.1595 \lxtwo
  \right]
  \nonumber\\&=&
  7.5018 \pm  0.0026
  -79.3384  \XLtwo
  +   187.0352  \XLtwotwo
  \nonumber\\&&\mbox{}
  + \XX \left[  0.0304 \pm  0.0022
  +  0.3770  \XLtwo
  +  1.2061  \XLtwotwo
  \right]
  \nonumber\\&&\mbox{}
  + \XXtwo \left[ -0.000393 \pm  0.000010
  -0.003743  \XLtwo
  -0.004326  \XLtwotwo
  -0.02711  \XLtwothree
  -0.01680  \XLtwofour
  \right]
  \nonumber\\&&\mbox{}
  + \XXthree  \left[ 
  - 0.0000055
  + 0.0000029 \XLtwo
  + 0.0000423  \XLtwotwo
  \right]
  \nonumber\\&=&
  \left[  115.1986 \pm  0.0026
  \right]
  + \XX \left[  1.6135 \pm  0.0022
  \right]
  + \XXtwo \left[ -0.052378 \pm  0.000010
  \right]
  \nonumber\\&&\mbox{} \qquad\qquad
  + \XXthree \left[  0.000040  \right]
  \nonumber\\&=&
  116.7598 \pm  0.0034
\label{eq::a8IVa0}\,,
\\
  A_2^{(8),\rm IV(a1)} &=&
   2.734 \pm  0.028
  \nonumber\\&&\mbox{}
  + \x \left[ -9.5571 
  \right]
  \nonumber\\&&\mbox{}
  + \xtwo \left[ -7.494 \pm  0.033
  -14.2010  \lx
  +  0.3559  \lxtwo
  \right]
  \nonumber\\&&\mbox{}
  + \xthree \left[  6.1668 
  -9.0654  \lx
  \right]
  \nonumber\\&=&
   2.734 \pm  0.028
  \nonumber\\&&\mbox{}
  + \XX \left[ -0.04622
  \right]
  \nonumber\\&&\mbox{}
  + \XXtwo \left[ -0.000175 \pm 0.000001
  +  0.001771  \XLtwo
  +  0.0002366  \XLtwotwo
  \right]
  \nonumber\\&&\mbox{}
  + \XXthree \left[ 0.000001 + 0.000005 \XLtwo  \right]
  \nonumber\\&=&
  \left[  2.734 \pm  0.028
  \right]
  + \XX \left[ -0.04622 
  \right]
  + \XXtwo \left[ 0.001832 \pm 0.000001  \right]
  + \XXthree \left[ 0.000006 \right]
  \nonumber\\&=&
   2.690 \pm  0.028
\label{eq::a8IVa1}\,,
\\
  A_2^{(8),\rm IV(a2)} &=&
  0.370 \pm  0.033 -0.7420 \lx
  \nonumber\\&&\mbox{}
  + \x \left[ 0  \right]
  \nonumber\\&&\mbox{}
  + \xtwo \left[  0.8628 \pm  0.002618
  \right]
  \nonumber\\&&\mbox{}
  + \xthree \left[ -2.6844 
  \right]
  \nonumber\\&=&
  + \xdummy \left[   0.370 \pm  0.033
  + 3.9561 \XLtwo
  \right]
  \nonumber\\&&\mbox{}
  + \XX \left[ 0  \right]
  \nonumber\\&&\mbox{}
  + \XXtwo \left[  0.00002018 \pm 0.00000006 \right]
  \nonumber\\&&\mbox{}
  + \XXthree \left[ -0.0000003  \right]
  \nonumber\\&=&
  + \xdummy \left[  4.326 \pm  0.03288
  \right]
  + \XX \left[ 0  \right]
  + \XXtwo \left[  0.00002018 \pm 0.00000006 \right]
  + \XXthree \left[ -0.0000003  \right]
  \nonumber\\&=&
  4.326 \pm  0.033
\label{eq::a8IVa2}\,,
\\
  A_2^{(8),\rm IV(b)} &=&
  27.395 \pm  0.014
  + ( 4.93482 \pm  0.00003) \lx
  \nonumber\\&&\mbox{}
  + \x \left[ -0.81 \pm  1.22
  +  59.0235 \lx
  \right]
  \nonumber\\&&\mbox{}
  + \xtwo \left[  142.5 \pm  7.6
  +  40.6546  \lx
  +  20.5582  \lxtwo
  -9.6167  \lxthree
  +  0.8333  \lxfour
  \right]
  \nonumber\\&&\mbox{}
  + \xthree \left[  62.11 \pm   2.89
  +  132.7421  \lx
  -40.9406  \lxtwo
  \right]
  \nonumber\\&=&
  27.395 \pm  0.014
  + (-26.3105 \pm 0.0002) \XLtwo
  \nonumber\\&&\mbox{}
  + \XX \left[ -0.0039 \pm  0.0059
  -1.5219  \XLtwo
  \right]
  \nonumber\\&&\mbox{}
  + \XXtwo \left[  0.003334 \pm  0.0001769
  -0.005070  \XLtwo
  +  0.01367  \XLtwotwo
  +  0.03409  \XLtwothree
  +  0.01575  \XLtwofour
  \right]
  \nonumber\\&&\mbox{}
  + \XXthree \left[  0.000007 \pm 0.
   -0.000080  \XLtwo
   -0.000132  \XLtwotwo
  \right]
  \nonumber\\&=&
   \xdummy \left[  1.084 \pm  0.014 \right]
  + \XX \left[ -1.5259 \pm  0.0059 \right]
  + \XXtwo \left[  0.06177 \pm  0.00018 \right]
  + \XXthree \left[ -0.0002047  \right]
  \nonumber\\&=&
  -0.380 \pm  0.016
\label{eq::a8IVb}\,,
\\
  A_2^{(8),\rm IV(c)} &=&
  -14.900 \pm  0.059  -3.2899  \lx
  \nonumber\\&&\mbox{}
  + \x \left[  65.4209
  \right]
  \nonumber\\&&\mbox{}
  + \xtwo \left[  33.61 \pm  9.96
  + ( 56.76 \pm 0.78) \lx
  + (-24.66 \pm 0.27) \lxtwo
\right.\nonumber\\&&\left.\qquad\qquad
  + ( -1.935 \pm 0.091) \lxthree
  \right]
  \nonumber\\&&\mbox{}
  + \xthree \left[  25.11 \pm  0.60
  +  107.4325 \lx
     -19.8610 \lxtwo
  \right]
  \nonumber\\&=&
  -14.9 \pm  0.05945 +  17.54 \XLtwo
  \nonumber\\&&\mbox{}
  + \XX \left[  0.3164 
  \right]
  \nonumber\\&&\mbox{}
  + \XXtwo \left[  0.00079 \pm  0.00023
  + (-0.007078 \pm 0.000097) \XLtwo
\right.\nonumber\\&&\left.\qquad\qquad
  + (-0.01640 \pm  0.00018) \XLtwotwo
  + ( 0.00686 \pm 0.00032) \XLtwothree
  \right]
  \nonumber\\&&\mbox{}
  + \XXthree \left[ 0.00000284 \pm 0.00000007 
  -0.00006479  \XLtwo
  -0.00006387  \XLtwotwo
  \right]
  \nonumber\\&=&
    \xdummy \left[  2.641 \pm  0.059  \right]
  + \XX \left[  0.3164   \right]
  + \XXtwo \left[ -0.01583 \pm 0.00045  \right]
  \nonumber\\&&\qquad\qquad
  + \XXthree \left[ -0.00012582 \pm 0.00000007 \right]
  \nonumber\\&=&
  2.941 \pm  0.059
\label{eq::a8IVc}  \,,
\end{eqnarray}
where the uncertainties origin from the numerical integration using
{\tt FIESTA}.
It is common to all five cases that the $x^3$ term only provides a negligible
contribution of at most 0.05\% [in the case of IV(b)]
which is much smaller than the
uncertainty estimate of the leading $x^0$ term.  In most cases 
the $x^2$ terms lead to contributions comparable to the numerical uncertainty
of the leading term. Together with the $x^3$ terms they confirm the good
convergence property of the asymptotic expansion.  It is nevertheless
interesting to note that in general the cubic and quartic logarithms of the
$x^2$ term lead to the largest numerical contributions, however, also the
quadratic $\log(x)$ terms are not negligible.

In the linear mass correction terms we observe logarithmic
contributions up to second order. They lead to significantly larger
numerical contributions than the linear logarithms which are in turn much
larger than the constant.

In the leading term only IV(a0) has a quadratic logarithm, one from each
electron loop. The $\log^2(x)$ term provides the numerical dominant contribution,
however, more than 40\% are canceled by the linear logarithm; the constant
term is an order of magnitude smaller. Linear logarithmic terms are also present
for IV(a2), IV(b) and IV(c). In the latter two cases strong cancellations
between the $\log(x)$ and the constant are observed, which 
leads to an interesting effect for IV(b): the leading term is 
smaller than the $m_e/m_\mu$-suppressed term. As a consequence the final
numerical result for IV(b) is quite small and has a big relative uncertainty.

\begin{table}[t]
  \begin{center}
  \begin{tabular}{l|r|r|c|c}
    $A_2^{(8)}(m_\mu / m_e)$ & 
    this work & \multicolumn{1}{c|}{\cite{Kinoshita:2004wi,Aoyama:2012wk}} & \cite{Calmet:1975tw} & \cite{Chlouber:1977dr} \\
    \hline
    IV(a0) & $116.76 \pm 0.02$ & $116.759183  \pm 0.000292$ & $111.1\pm8.1$ & $117.4\pm0.5$ \\
    IV(a1) & $2.69 \pm 0.14$   & $2.697443 \pm 0.000142$ && \\
    IV(a2) & $4.33 \pm 0.17$  & $4.328885 \pm 0.000293$ && \\
    IV(a) & $123.78\pm 0.22$ & $123.78551\hphantom{2} \pm 0.00044\hphantom{2}$ & &\\
    IV(b) & $-0.38 \pm 0.08$ & $-0.4170\hphantom{22}   \pm 0.0037\hphantom{22}$ & & \\
    IV(c) & $2.94 \pm 0.30$  & $2.9072\hphantom{22}    \pm 0.0044\hphantom{22}$ & & \\
  \end{tabular}
  \caption{\label{tab::a8}Summary of the final results for the individual
    four-loop light-by-light-type contributions and their comparison with 
    results from the literature. Note that the uncertainties given in the
    second column are obtained from Eqs.~(\ref{eq::a8IVa0})
    to~(\ref{eq::a8IVc}) after multiplication by five.}
  \end{center}
\end{table}

In Table~\ref{tab::a8} we summarize our findings for IV(a0), IV(a1), IV(a2),
IV(b) and IV(c) and compare with the literature.
To be on the conservative side we multiply the Monte-Carlo uncertainty
from Eqs.~(\ref{eq::a8IVa0}) to~(\ref{eq::a8IVc}) by a factor of five.
For all classes under consideration we find good agreement with previous
results. For the dominant contribution IV(a) we find excellent agreement.

Although our numerical precision can not compete with the one of
Ref.~\cite{Aoyama:2012wk}, let us note that our computation procedure is
completely different from the numerical method used
in~\cite{Aoyama:2012wk}. As our calculation is a second evaluation of the 
complete light-by-light contribution to $A_2^{(8)}(m_\mu / m_e)$, our result is
an important check of the existing value. Furthermore, since the asymptotic
expansion and the reduction to master integrals were done analytically, it is
possible to systematically improve the precision by evaluating more and
more master integrals analytically.


\section{\label{sec::concl}Conclusions}

We have computed the four-loop QED corrections to the muon magnetic moment
induced by the light-by-light-type Feynman diagrams involving a closed
electron loop. This is a finite and gauge invariant subset, which provides
about 95\% of the total four-loop contribution.  Our results for the
individual sub-classes IV(a), IV(b) and IV(c) agree with the
literature~\cite{Aoyama:2012wk}.

We want to stress that our approach is completely different from the one of
Ref.~\cite{Aoyama:2012wk} and thus should be considered as an independent cross
check.  We perform an analytic reduction of all occurring integrals to a small
set of master integrals which are then computed using analytical or numerical
methods. In particular all counterterm contributions are available
analytically. On the other hand, in Ref.~\cite{Aoyama:2012wk} infrared and
ultraviolet finite multi-dimensional integrands for the individual sub-classes
of $a_\mu^{(8)}$ are constructed which are then integrated numerically.

Our final results can be found in Table~\ref{tab::a8}.  Although our numerical
uncertainty, which amounts to approximately $0.4 \times (\alpha/\pi)^4 \approx
1.2 \times 10^{-11}$, is larger than the one of Ref.~\cite{Aoyama:2012wk}, it
is still much smaller than the difference between the theory prediction and
experimental result of $a_\mu$ [cf. Eq.~(\ref{eq::amu_diff})].
We also want to note that our result can be systematically improved by
evaluating more and more master integrals analytically.


\section*{Acknowledgments}

We would like to thank M.~Nio for useful communications.
We thank the High Performance Computing Center
Stuttgart (HLRS) and the Supercomputing Center of Lomonosov Moscow
State University~\cite{LMSU} for providing computing time used for the
numerical computations with {\tt FIESTA}.  P.M was supported in part
by the EU Network HIGGSTOOLS PITN-GA-2012-316704.  This work was
supported by the DFG through the SFB/TR~9 ``Computational Particle
Physics''.  The work of V.S. was supported by the Alexander von
Humboldt Foundation (Humboldt Forschungspreis).



\section*{Appendix: Selected analytic three- and four-loop results for $a_\mu$}

In this appendix we provide analytic results for the three-loop light-by-light
contribution $A_{2, \rm lbl}^{(6)}(m_\mu / m_e)$ and for some of the
four-loop contributions.

The analytic version of Eq.~(\ref{eq::a6}) reads
\begin{align}
  A_{2, \rm lbl}^{(6)}(m_\mu / m_e) 
=\ & x^0\ \left( \frac{2}{3} - \frac{10 \pi^2}{3} + \frac{59 \pi^4}{270} - 3 \zeta_3 - \frac{2 \pi^2}{3}\ \ell_x \right) \nonumber\\
&+x \left( \frac{424 \pi^2}{9} - \frac{196 \ln(2) \pi^2}{3} - \frac{4 \pi^2}{3}\ \ell_x \right) \nonumber\\
&+x^2\ \Bigg[ -\frac{283}{12} + \frac{25 \pi^2}{18} - \frac{61 \pi^4}{270} + 3 \zeta_3 + \frac{4 \pi^2 \zeta_3}{3} \nonumber\\
& + \left( \frac{61}{3} - \frac{32 \pi^2}{9} + \frac{16 \pi^4}{135} + 4 \zeta_3 \right)\ \ell_x + \left( \frac{\pi^2}{9} - \frac{20}{3} \right)\ \ell_x^2 + \frac{2}{3}\ \ell_x^3 \Bigg] \nonumber\\
&+x^3\ \left( -\frac{11 \pi^2}{9} - \frac{10 \pi^2}{9}\ \ell_x \right) \nonumber\\
&+x^4\ \left[ \frac{13283}{2592} + \frac{191 \pi^2}{216} + \frac{\zeta_3}{2} + \left( - \frac{517}{108} - \frac{13 \pi^2}{9} \right)\ \ell_x + \frac{41}{18}\ \ell_x^2 - \frac{7}{9}\ \ell_x^3 \right]
  \,.
\end{align}
where $\zeta_n$ is Riemann's zeta function evaluated at $n$.

At four-loop order one can obtain analytic results for the coefficients of
$\lx^n$ ($n\ge1$) for IV(a0), IV(a1) and IV(a2), for the $x^1$ and $x^3$ terms
of IV(b) and for the $x^0$, $x^1$ and $x^3$ terms of IV(c). These
contributions are given by
\begin{align}
A_2^\text{(8),IV(a0)}|_{\ell_x} =\ & \left( -\frac{4}{3} + \frac{8 \pi^2}{3} - \frac{8 \pi^4}{45} + 6 \zeta_3 \right)\ \ell_x + \frac{2 \pi^2}{3}\ \ell_x^2 \nonumber\\
&+x \left( \frac{-40 \pi^2}{27}\ \ell_x + \frac{8 \pi^2}{9}\ \ell_x^2 \right) \nonumber\\
&+x^2\ \Bigg[ \left( \frac{47}{3} + 6 \pi^2 - \frac{98 \pi^4}{405} - 10 \zeta_3 - \frac{8 \pi^2 \zeta_3}{27} - \frac{50 \zeta_5}{9} \right)\ \ell_x \nonumber\\
&+ \left( -\frac{220}{9} + \frac{32 \pi^2}{9} - \frac{16 \pi^4}{135} - \frac{14 \zeta_3}{3} \right)\ \ell_x^2 + \left( \frac{82}{9} - \frac{4 \pi^2}{27} \right)\ \ell_x^3 -\frac{8}{9}\ \ell_x^4 \Bigg] \nonumber\\
&+x^3\ \left( \frac{-40 \pi^2}{81}\ \ell_x + \frac{4 \pi^2}{3}\ \ell_x^2 \right) \,,\\
A_2^\text{(8),IV(a1)}|_{\ell_x} =\ & x^2\ \left( \frac{137}{27} - \frac{12629 \pi^2}{9720} - \frac{2 \ln(2) \pi^2}{3} - \frac{31 \pi^4}{405} + \frac{125 \zeta_3}{27} \right)\ \ell_x \nonumber\\
&+ x^2\ \left( \frac{\pi^2}{9} - \frac{20}{27} \right)\ \ell_x^2 - x^3\ \frac{124 \pi^2}{135}\ \ell_x\,,\\
A_2^\text{(8),IV(a2)}|_{\ell_x} =\ & \bigg( -\frac{10}{9} - 32 a_4 - \frac{4 \ln^4(2)}{3} - \frac{931 \pi^2}{27} + 48 \ln(2) \pi^2 \nonumber\\
&+ \frac{4 \ln^2(2) \pi^2}{3} + \frac{41 \pi^4}{270} + \frac{8 \zeta_3}{3} + \frac{5 \pi^2 \zeta_3}{9} - \frac{5 \zeta_5}{3} \bigg)\ \ell_x \,,\\
A_2^\text{(8),IV(b)}|_{x^1,\ell_x} =\ & x \left(3 \pi^3 - \frac{31 \pi^2}{9}\right)\ \ell_x \,,\\
A_2^\text{(8),IV(b)}|_{x^3,\ell_x} =\ & x^3\left[ \left(\frac{137 \pi^2}{4050} + \frac{1153 \pi^3}{270}\right)\ \ell_x - \frac{112 \pi^2}{27}\ \ell_x^2 \right]\,,\\
A_2^\text{(8),IV(c)}|_{x^0,\ell_x} =\ & -\frac{\pi^2}{3}\ \ell_x \,,\\
A_2^\text{(8),IV(c)}|_{x^1,\ell_x} =\ & 0 \,,\\ 
A_2^\text{(8),IV(c)}|_{x^3,\ell_x} =\ & x^3 \left[ \frac{2939 \pi^2}{270}\
  \ell_x - \frac{163 \pi^2}{81}\ \ell_x^2 \right] \,,
\end{align}
with $a_4=\mbox{Li}_4(1/2)$.




\end{document}